\documentclass [11pt,a4paper] {article}

\usepackage[cp1252]{inputenc}

\usepackage{amssymb}
\usepackage{amsmath}
\usepackage{amsfonts,amssymb}

\usepackage[dvips]{graphicx}

\usepackage{bbm}

\DeclareMathAlphabet{\mathpzc}{OT1}{pzc}{m}{it}

\newcommand*\rfrac[2]{{}^{#1}\!\!/\!_{#2}}

\begin{document}

\title{Truth values of quantum phenomena}

\author{Arkady Bolotin\footnote{$Email: arkadyv@bgu.ac.il$\vspace{5pt}} \\ \textit{Ben-Gurion University of the Negev, Beersheba (Israel)}}

\maketitle

\begin{abstract}\noindent In the paper, the idea of describing not-yet-verified properties of quantum objects with logical many-valuedness is scrutinized. As it is argued, to promote such an idea, the following two foundational problems of many-valued quantum logic must be decided: the problem of choosing a proper system of many-valued logic and the problem of the emergence of bivalence from logical many-valuedness. Difficulties accompanying solutions of these problems are discussed.\\

\end{abstract}

\section{Introducing logical many-valuedness in quantum mechanics}\label{Introduction}  

\noindent The argument claiming that quantum theory could not be comprehended on the grounds of classical two-valued logic is rather straightforward and goes like this.\\

\noindent Let us consider a typical quantum interference experiment where a quantum particle being released from a source is absorbed by a screen after passing through a two-slit barrier
$\,$\footnote{\label{f1}In the present paper, rather than being strictly restricted to spatially arranged slits, quantum interference is considered generally for any set of perfectly distinguishable alternatives.\vspace{5pt}}. Suppose that immediately behind that barrier are placed two which-way detectors able to verify (e.g., by way of clicking) the particle's passage through a corresponding slit. Let $X_1$ denote the proposition of the click of the detector placed behind slit 1 such that $X_1$ is true (denoted by ``1'') if the detector clicks and $X_1$ is false (denoted by ``0'') if the detector does not. Let $X_2$ in an analogous manner denote the proposition of the signal from the detector placed behind slit 2.\\

\noindent Assume that the propositions $X_1$ and $X_2$ are in possession of \textit{not-yet-verified truth values} -- i.e., ones existing before the detectors can click -- that are merely  revealed by the act of verification of the particle's passage.\\

\noindent Within the given assumption, let us accept that such values of the propositions $X_1$ and $X_2$ are \textit{either 1 or 0}. Accordingly, exclusive disjunction on these values of $X_1$ and $X_2$ can be decided by\smallskip

\begin{equation} \label{1} 
      X_{12}
      \equiv
      X_1 \,\underline{\lor}\, X_2
      \equiv
      \left(
         \bigvee_{i=1}^{2}X_i
      \right)
      \land
      \neg 
      \left(
         \bigwedge_{i=1}^{2}X_i
      \right)
      \;\;\;\;  .
\end{equation}
\smallskip

\noindent Suppose that the not-yet-verified truth value of the compound proposition $X_{12}$ is true. Provided that $\mathbb{P}$ is the probability function mapping propositions $Y,Z,\dots$ to the unit interval $[0,1]$ such that $\mathbb{P}[Y]=1$ if $Y$ is true, $\mathbb{P}[Y]=0$ if $Y$ is false, and $\mathbb{P}[Y \lor Z]=P[Y]+P[Z]-P[Y \land Z]$, the probability of finding the particle at a certain region $R$ on the screen would be then given by the sum of the patterns $\mathbb{P}[R|X_i]$ emerging in setups with a one-slit barrier, namely, $\mathbb{P}[R|X_1 \lor X_2] = \rfrac{1}{2} (\mathbb{P}[R|X_1] + \mathbb{P}[R|X_2])$ (on condition that $\mathbb{P}[X_1] = \mathbb{P}[X_2]$). It would mean that in the said case, quantum interference would be nonexistent even with none of the detectors present at the slits.\\

\noindent So, by contrast, let us suppose that the not-yet-verified truth value of $X_{12}$ is false. But then -- in contradiction to the quantum collapse postulate -- one would find that it is not true that exactly one detector will signal if the particle's passage through the two-slit barrier is observed.\\

\noindent Thus, for the assumption of the not-yet-verified truth values of the propositions $X_i$ to be consistent with the occurrence of quantum interference and quantum collapse, these truth-values must be \textit{neither 1 nor 0}. Such could be only if prior to their verification $X_{i}$ does not obey the principle ``\textit{a proposition is either true or false}'', i.e., the principle of bivalence
$\,$\footnote{\label{f2}Obviously, it is possible to avoid this conclusion merely by accepting \textit{nonlocal realism} (i.e., an interpretation of quantum theory in terms of `hidden variables' such as Bohmian mechanics \cite{Bohm1,Bohm2,Bohm3}). But in doing so one would confront with additional deficiencies that plague the `hidden variables' approach (the analysis of those deficiencies can be found, e.g., in \cite{Rainer,Jung}).\vspace{5pt}}.\\

\noindent From the violation of this principle, one can infer that results of future non-certain events can be described using \textit{many-valued logics}. For example, consider a 3-valued logic $\{0,\rfrac{1}{2},1\}$ which includes only one additional truth value $\rfrac{1}{2}$ besides the classical ones 0 and 1
$\,$\footnote{\label{f3}That might be such three-valued logical systems as the Kleene (strong) logic $K_3$ or the 3-valued {\L}ukasiewicz system \cite{Gottwald,Miller}.\vspace{5pt}}. If not-yet-verified truth values of the propositions $X_i$ are both $\rfrac{1}{2}$, where the truth value $\rfrac{1}{2}$ is interpreted as ``possible'' and the valuations $\neg \rfrac{1}{2} = \rfrac{1}{2}$, $\rfrac{1}{2} \lor \rfrac{1}{2} =\rfrac{1}{2}$, $\rfrac{1}{2} \land \rfrac{1}{2} = \rfrac{1}{2}$ hold, then the truth value of the compound proposition $X_{12}$ can be ascertained as $(\rfrac{1}{2} \lor \rfrac{1}{2}) \land \neg (\rfrac{1}{2} \land \rfrac{1}{2}) =\rfrac{1}{2}$, i.e., also ``possible''. In this case, before the verification one can only assert that the both statements -- \textit{the particle passes through exactly one slit} and \textit{the particle passes through more than one slit} -- are possible.\\

\noindent Attractive as it might seem, the idea of describing not-yet-verified properties of quantum objects with many-valued logics is burdened with two foundational problems.\\

\noindent The first is \textit{the problem of choosing a proper system of many-valued logic}. It concerns with the following question: Because there are infinitely many systems of many-valued logic, which of them should be chosen for the quantum mechanical description? How can a specific system be decided on to avoid a charge of arbitrariness?\\

\noindent Notwithstanding the significance of the first problem, the next one seems even more serious: When the act of the verification is finished in the macrophysical domain (and so the path the particle has taken is known with certainty), the propositions $X_i$ conform to the principle of bivalence. So, unless one can demonstrate that a many-valued logic has a proper application to the objects of everyday macrophysical experience (and for this reason our logic needs revising), the following question must be answered: \textit{How do not-yet-verified multivalent truth values become classical bivalent truth values?}\\

\noindent The discussion of these foundational problems of many-valued quantum logic is offered in the present paper.\\

\section{Preliminaries}\label{Preliminaries}  

\noindent We will start by briefly introducing a few necessary preliminaries.\\

\noindent Let us consider a complete lattice $\mathcal{L}=(L,\sqcup,\sqcap)$ containing any partially ordered set $L$ where each two-element subset $\{y,z\} \subseteq L$ has a join (i.e., a least upper bound) and a meet (i.e., a greatest lower bound) defined by $y \sqcup z \equiv l.u.b.(y,z)$ and $y\sqcap z \equiv g.l.b.(y,z)$, correspondingly. In addition to the binary operations $\sqcup$ and $\sqcap$, let the lattice $\mathcal{L}$ contain a unary operation $\sim$ defined in a manner that $L$ is closed under this operation and $\sim$ is an involution, explicitly, ${\sim}y \in L$ if $y \in L$ and ${\sim}({\sim}y)=y$.\\

\noindent Let\smallskip

\begin{equation} \label{2} 
   \S^{\mathcal{V}}_{N}(\diamond)
   =
   {[\![ \diamond ]\!]}_v
   \;\;\;\;  ,
\end{equation}
\smallskip

\noindent where the symbol $\diamond$ can be replaced by any proposition associated with a property of a physical system, refer to a valuation, i.e., a mapping from a set of propositions being as stated denoted by $S=\{\diamond\}$ to a set $\mathcal{V}_N = \{ \mathfrak{v} \}$ where $\mathfrak{v}$ are truth-values ranging from 0 to 1 and $N$ is the cardinality of the set $\{ \mathfrak{v} \}$:\smallskip

\begin{equation} \label{3} 
   \S^{\mathcal{V}}_{N}
   :
   S \to \mathcal{V}_N
   \;\;\;\;  .
\end{equation}
\smallskip

\noindent At the same time, let us assume a homomorphism $f:L \to S$ such that there is a truth-function $v$ that maps each lattice element denoted by the symbol $\ast$ to the truth value of the corresponding proposition, namely,\smallskip

\begin{equation} \label{4} 
   v(\ast) = {[\![ \diamond ]\!]}_v
   \;\;\;\;  ,
\end{equation}
\smallskip

\noindent basing on the following principles:\smallskip

\begin{equation} \label{5} 
   v(y) = 0  \quad \text{if} \quad y = 0_L
   \;\;\;\;  ,
\end{equation}

\begin{equation} \label{6} 
   v(y) = 1  \quad \text{if} \quad y = 1_L
   \;\;\;\;  ,
\end{equation}
\smallskip

\noindent where $0_L$ and $1_L$ are the least and the greatest elements of the lattice, correspondingly. These principles imply that the least and the greatest lattice elements are identified with always false and always true propositions.\\

\noindent Let the following valuation apply for the negation of a proposition $Y$:\smallskip

\begin{equation} \label{7} 
    v({\sim}y)
    =
   {[\![ \neg Y ]\!]}_v
   \;\;\;\;  .
\end{equation}
\smallskip

\noindent On the other hand, the valuation ${[\![ \neg Y ]\!]}_v$ can be decided through the truth degree function $F_{\neg}$ of negation, that is,

\begin{equation} \label{8} 
   {[\![ \neg Y ]\!]}_v
   =
   F_{\neg} \!
   \left(
       {[\![Y]\!]}_v
   \right)
   \;\;\;\;  .
\end{equation}
\smallskip

\noindent As stated by \cite{Gottwald}, the most basic example for $F_{\neg}$ is $1 - {[\![Y]\!]}_v$ (called \textit{{\L}ukasiewicz negation}). To meet this version of negation, let us assume\smallskip

\begin{equation} \label{9} 
   v({\sim}y)
   =
   1 - v(y)
   \;\;\;\;  .
\end{equation}
\smallskip

\noindent Also, let the following valuational axioms apply for the binary operations on lattice elements $y$ and $z$ related to \textit{simultaneously verifiable} propositions $Y$ and $Z$:\smallskip

\begin{equation} \label{10} 
    v(y \sqcup z)
    =
   {[\![ Y \!\lor\! Z ]\!]}_v
   \;\;\;\;  ,
\end{equation}

\begin{equation} \label{11} 
    v(y \sqcap z)
    =
   {[\![ Y \!\land\! Z ]\!]}_v
   \;\;\;\;  .
\end{equation}
\smallskip

\noindent From another side, the truth values of disjunction and conjunction on the values of the propositions $Y$ and $Z$ can be decided by

\begin{equation} \label{12} 
   {[\![ Y \!\lor\! Z ]\!]}_v
   =
   F_{\lor} \!
   \left(
       {[\![Y]\!]}_v,  {[\![Z]\!]}_v
   \right)
   \;\;\;\;  ,
\end{equation}

\begin{equation} \label{13} 
   {[\![ Y \!\land\! Z ]\!]}_v
   =
   F_{\land} \!
   \left(
       {[\![Y]\!]}_v,  {[\![Z]\!]}_v
   \right)
   \;\;\;\;  ,
\end{equation}
\smallskip

\noindent where $F_{\lor}$ and $F_{\land}$ are the truth degree functions of the corresponding logical connectives.
$\,$\footnote{\label{f4} In \cite{Pykacz17}, the relation between the functions $v(y \sqcup z)$ and $F_{\lor} \!\left({[\![Y]\!]}_v, {[\![Z]\!]}_v\right)$ as well as $v(y \sqcap z)$ and $F_{\land} \!\left({[\![Y]\!]}_v, {[\![Z]\!]}_v\right)$ is studied to examine whether {\L}ukasiewicz operations can also be used to model conjunctions and disjunctions. As it is stated in the paper, {\L}ukasiewicz disjunction and conjunction coincide with the truth-functions of joins and meets, namely, $v(y \sqcup z)=\min{\{v(y)+v(z),1\}}$ and $v(y \sqcap z)=\max{\{v(y)+v(z)-1,0\}}$, whenever these {\L}ukasiewicz connectives can be defined.\vspace{5pt}}\\

\noindent As it has been already noted in the Section \ref{Introduction}, were the simultaneously verifiable propositions $X_i$ to possess not-yet-verified truth-values consistent with the occurrence of quantum interference and quantum collapse, they would have to meet the requirement $\{v(x_i)\} \neq \{0,1\}$, where $x_i$ are the lattice elements attributed to the propositions $X_i$ such that $v(x_i) = {[\![X_i]\!]}_v$. But together with that, after the verification, the truth-values of $X_i$ would have to conform to bivalence, i.e., $\{v(x_i)\} = \{0,1\}$.\\

\noindent Consequently, the question is, how to combine in one theory those two mutually exclusive conditions on $\{v(x_i)\}$?\\

\section{Many-valued quantum logic}\label{MVQL}  

\noindent A solution to this problem can be motivated by recalling that the total of all the individual probabilities equals 1, so, when one of the probabilities turns into 1, all the others become 0. Let us add some details to this idea
$\,$\footnote{\label{f5} In fact, this idea -- called \textit{many-valued quantum logic} or \textit{fuzzy quantum logic} -- has already been developed in a series of papers \cite{Pykacz94,Pykacz95,Pykacz00,Pykacz10,Pykacz11,Pykacz15}; however, for the aim of this discussion, it is not necessary to follow those papers precisely. Also, for the discussion it is immaterial to present in its entirety the generally accepted interpretation of the elements of a quantum logic -- an interested reader can be referred to any textbook on quantum logic: see, for example, \cite{Beltrametti} or \cite{Pitowsky}.\vspace{5pt}}.\\

\noindent Assume that there is a correspondence (homomorphism) between a lattice $\mathcal{L}=(L,\sqcup,\sqcap)$ and a family $L(\mathcal{H})$ containing all closed, ordered by the subset relation  subspaces of a (separable) Hilbert space $\mathcal{H}$ associated with a physical system under investigation. Explicitly, the inclusion relation between the lattice elements corresponds to the subset relation between the closed subspaces, the lattice meet $\sqcap$ corresponds to the intersection $\cap$ of those subspaces and the lattice join $\sqcup$ is the closed span of their union $\cup$, the least element of the lattice is the $\{0\}$ subspace and the greatest element of the lattice is the identical subspace $\mathcal{H}$.\\

\noindent Consider self-adjoint projection operators $\hat{P}_{\diamond}$ on $\mathcal{H}$ that represent propositions $\diamond$ declaring that the system possesses experimentally verifiable properties (such as a path that the particle takes getting through the barrier). Since each projection operator $\hat{P}_{\diamond}$ leaves invariant any vector lying in its range, $\mathrm{ran}(\hat{P}_{\diamond})$, and annihilates any vector lying in its null space, $\mathrm{ker}(\hat{P}_{\diamond})$, giving in this manner a decomposition of $\mathcal{H}$ into two complementary closed subspaces, namely, $\mathcal{H}=\mathrm{ran}(\hat{P}_{\diamond}) \oplus \mathrm{ker}(\hat{P}_{\diamond})$, there is a one-one correspondence between the subspaces $\mathrm{ran}(\hat{P}_{\diamond})$ and projection operators $\hat{P}_{\diamond}$. Therefore, one can consider $\hat{P}_{\diamond}$ as the elements of $L(\mathcal{H})$.\\

\noindent In view of the homomorphism between $\mathcal{L}=(L,\sqcup,\sqcap)$ and $L(\mathcal{H})$, the principles of calculus of truth values presented above are expected to survive the passage to compatible elements of $L(\mathcal{H})$. For this reason, one can put forward that\smallskip

\begin{equation} \label{14} 
    v(\hat{P}_x)
    =
   {[\![ X ]\!]}_v
   \;\;\;\;  ,
\end{equation}
\smallskip

\noindent where $v(\hat{P}_x)$ is a truth-function value at projection operator $\hat{P}_x$ which corresponds to a truth value of proposition $X$.\\

\noindent Suppose that the state of the system is characterized by the unit vector $|\Psi\rangle$. It is not difficult to see that the truth-value of the proposition $X$ will coincide with the eigenvalue of the projection operator $\hat{P}_x$, namely, $\{v(\hat{P}_x)\} = \{0,1\}$, if and only if the given vector $|\Psi\rangle$ lies in either $\mathrm{ran}(\hat{P}_x)$ or $\mathrm{ker}(\hat{P}_x)$.\\ 

\noindent However, if $|\Psi\rangle$ is such a unit vector in $\mathcal{H}$ that $|\Psi\rangle \not\in \mathrm{ran}(\hat{P}_x)$ as well as $|\Psi\rangle \not\in \mathrm{ker}(\hat{P}_x)$, one can only get $\{\langle\Psi|\hat{P}_x|\Psi\rangle\} = \{\bar{x} \in \mathbb{R}\,|\, 0 < \bar{x} < 1 \}$, where $\bar{x}$ is the expected value of the observable $x$ corresponding to the operator $\hat{P}_x$. Consistent with the orthodox quantum theory, the value $\langle\Psi|\hat{P}_x|\Psi\rangle$ can be interpreted as the probability that the measurement of the observable $x$ will produce the “affirmative” answer 1 in the state of the system given by $|\Psi\rangle$, i.e.,\smallskip

\begin{equation} \label{15} 
   \mathbb{P}[x=1]
   \equiv
   \langle\Psi|\hat{P}_x|\Psi\rangle
   \;\;\;\;  .
\end{equation}
\smallskip

\noindent But, as stated by the idea being discussed here, the value $\langle\Psi|\hat{P}_x|\Psi\rangle$ should be also regarded as the not-yet-verified truth value of the proposition $X$, namely,\smallskip

\begin{equation} \label{16} 
   v(\hat{P}_x)
   =
   \langle\Psi|\hat{P}_x|\Psi\rangle
   \;\;\;\;  .
\end{equation}
\smallskip

\noindent Accordingly, if $|\Psi\rangle$ is any unit vector in $\mathcal{H}$, then $\{v(\hat{P}_x)\} = \{\bar{x} \in \mathbb{R}\,|\, 0 \le \bar{x} \le 1 \}$ which can be interpreted as a generalization of the Boolean domain $\mathcal{V}_2 =\{0,1\}$. In this fashion, the value $\langle\Psi|\hat{P}_x|\Psi\rangle$ represents the degree to which the proposition $X$ is true prior to the verification (i.e., before the experiment designed to verify the affirmative answer can be completed)
$\,$\footnote{\label{f6}At the same time, $v(\sim\!\hat{P}_x) = 1-\langle\Psi|\hat{P}_x|\Psi\rangle$ represents the degree to which the not-yet-verified truth value of the proposition $X$ is not true (that is, the degree to which the system does not possess the mentioned property prior to the verification).\vspace{5pt}}.\\

\section{Truth-values vs. probabilities}\label{TVvsProb}  

\noindent Let us analyze the appropriateness of the hypothesis (\ref{16}).\\

\noindent Firstly, consider the rationale behind it. Suppose that a quantum system is prepared in a pure normalized state $|\Psi\rangle$ that lies in the range of the projection operator $\hat{P}_x$. Being in the state $|\Psi\rangle$ is subject to the assumption that the truth-value function $v$ must assign the truth value 1 to the proposition $X$ and, in consequence, to the operator $\hat{P}_x$, namely, $|\Psi\rangle \in \mathrm{ran}(\hat{P}_x)  \implies v(\hat{P}_x) = {[\![ X ]\!]}_v =1$, since $\hat{P}_x |\Psi\rangle =1 \cdot|\Psi\rangle$. But what is more, in this case, the affirmative answer for the measurement of the observable $x$ will have probability 1 since $\langle\Psi|\hat{P}_x|\Psi\rangle = 1$. In an analogous manner, if the system is prepared in a pure state $|\Psi\rangle$ lying in the null space of the projection operator $\hat{P}_x$, then the function $v$ must assign the truth value 0 to $\hat{P}_x$, namely, $|\Psi\rangle \in \mathrm{ker}(\hat{P}_x)  \implies v(\hat{P}_x) = {[\![ X ]\!]}_v =0$, since $\hat{P}_x |\Psi\rangle =0 \cdot|\Psi\rangle$. In that case, the probability of the affirmative answer must be 0 as $\langle\Psi|\hat{P}_x|\Psi\rangle = 0$. From here on can infer that the truth-function $v(\hat{P}_x)$ and the probability-function $\mathbb{P}[x=1]$ agree if $|\Psi\rangle \in \mathrm{ran}(\hat{P}_x)$ or $|\Psi\rangle \in \mathrm{ker}(\hat{P}_x)$.\\

\noindent It is tempting to conclude that the agreement\smallskip

\begin{equation} \label{17} 
   v(\hat{P}_x)
   =
   \mathbb{P}[x=1]
   \;\;\;\;   
\end{equation}
\smallskip

\noindent holds even in the case where $|\Psi\rangle \notin \mathrm{ran}(\hat{P}_x)$ and $|\Psi\rangle \notin \mathrm{ker}(\hat{P}_x)$. Sadly, such a conclusion is open to some considerable objections.\\

\noindent First, the equality (\ref{17}) is too strong from the mathematical point of view. Namely, the fact that two outputs of the functions $v(\hat{P}_x)$ and $\mathbb{P}[x=1]$ coincide does not mean that these functions have the same codomain. The failure of bivalence $\{v(\hat{P}_x)\} \neq \{0,1\}$ only suggests that the set of all permitted outputs to the truth-function $v(\hat{P}_x)$ may contains more than two values. That is, $v(\hat{P}_x)$ might be infinite-valued and yet different from $\mathbb{P}[x=1]$ in any state $|\Psi\rangle$ where $|\Psi\rangle \notin \mathrm{ran}(\hat{P}_x)$ and $|\Psi\rangle \notin \mathrm{ker}(\hat{P}_x)$.\\

\noindent Second, from the conceptual point of view, the equality (\ref{17}) is strong as well given that it is not \textit{conceptually neutral}. That is, the pertinence of this equality to quantum theory strongly depends on the interpretation of the state vector $|\Psi\rangle$. This means that the equality (\ref{17}) can be applicable only if the vector $|\Psi\rangle$ posits the ``true states of reality'', i.e., the ontic states of the quantum system.\\

\noindent E.g., in the Bayesian approach to quantum mechanics, probabilities -- and thus the state vector $|\Psi\rangle$ -- represent an agent's degrees of belief, rather than corresponding to objective properties of physical systems \cite{Caves}. As a result, within the Bayesian approach the equality (\ref{17}) would not be right since its left-hand side would be \textit{objective} while its right-hand side would be \textit{subjective}. In more detail, gathering data allows the agents to update their probability assignments (by using Bayes' theorem); so, the probability $\mathbb{P}[x=1]$ always depends on the agents' prior probabilities as well as on the data and therefore can be different for agents in possession of the same data.
$\,$\footnote{\label{f7}As stated by Bayesian approach to probability theory, probabilities are degrees of belief, not facts. Probabilities cannot be derived from facts alone. Two agents who agree on the facts can legitimately assign different prior probabilities. In this sense, probabilities are not objective, but subjective (see, e.g., \cite{Morgan,Savage,Finetti,Bernardo}).\vspace{5pt}} On the other hand, the proposition $X$ is the statement that in the measurement of the observable $x$ the outcome 1 occurs. Accordingly, the truth value of the proposition $X$ is a fact for any agent.\\

\noindent Finally, the equality (\ref{17}) gives rise to the problem of \textit{the emergence of bivalence from many-valued logics}. Indeed, if this equality holds as a general principle and hence the logic underpinning the reality is infinite-valued, then the question is, how does a two-valued semantics emerge from an infinite-valued semantics during the process of verification?\\

\noindent To describe the emergence of the logical bivalent limit, one can use clues suggested in the paper \cite{Losada}.\\

\noindent Consider a quantum system and a set $\mathcal{O}$ of projection operators on the Hilbert space related to the states for the system, namely, $\mathcal{O} = \{\hat{P}_{q\alpha},\hat{P}_{r\beta}\}$, where $\alpha = \{1,\dots,n\}$ and $\beta = \{1,\dots,m\}$, such that some projection operators of $\mathcal{O}$ are incompatible, that is, $[\hat{P}_{q\alpha},\hat{P}_{r\beta}] \equiv \hat{P}_{q\alpha} \hat{P}_{r\beta} - \hat{P}_{r\beta}\hat{P}_{q\alpha} \neq 0$. The incompatibility of the projection operators $\hat{P}_{q\alpha}$ and $\hat{P}_{r\beta}$ means that if the system is prepared, say, in the state $|{\Psi}_q\rangle$ where all the propositions $Q_{\alpha}$ comes out bivalent, then the truth values of the propositions $R_{\beta}$ cannot be two-valued in $|{\Psi}_q\rangle$: To be exact, $|{\Psi}_q\rangle \notin \mathrm{ran}(\hat{P}_{r\beta})$, so ${[\![ R_{\beta} ]\!]}_v \neq 1$; also, $|{\Psi}_q\rangle \notin \mathrm{ker}(\hat{P}_{r\beta})$, thus ${[\![ R_{\beta} ]\!]}_v \neq 0$.\\

\noindent Clearly, had the commutator $[\hat{P}_{q\alpha},\hat{P}_{r\beta}]$ been equal to 0, the propositions $Q_{\alpha}$ and $R_{\beta}$ would have become bivalent in the prepared state $|{\Psi}_q\rangle$. In view of that, the logical bivalent limit can be understood as a deformation of a non-commutative algebra and a limit $\hbar \to 0$.\\ 

\noindent Explicitly, assume that the noncommutative observables $q$ and $r$ have discrete spectrums $\{q_\alpha\}$ and $\{r_\beta\}$ such that for their operators $\hat{q}$ and $\hat{r}$ one can write $\hat{q}=\sum_{\alpha}^{n}q_\alpha \hat{P}_{q\alpha}$ and $\hat{r}=\sum_{\beta}^{m}r_\beta \hat{P}_{r\beta}$. Along these lines, in the limit $\hbar \to 0$ the commutation relation between the operators $\hat{q}$ and $\hat{r}$ can be presented in the following form\\

\begin{equation} \label{18} 
   [\hat{q},\hat{r}]
   \neq
   0
   \;\;
   \stackrel{\!\!\!\! \hbar \to 0}{\longrightarrow}
   \;\;
   \sum_{\alpha}^{n}
      \sum_{\beta}^{m}
         q_\alpha\ \!\! r_\beta
         [\hat{P}_{q\alpha},\hat{P}_{r\beta}]
   =
   i\hbar
   \{ q,r \}
   +
   O({\hbar}^2)
   \;\;\;\;  ,
\end{equation}
\smallskip

\noindent where $\{ q,r \}$ denotes the classical counterpart of the commutator $ [\hat{q},\hat{r}]$.\\

\noindent At this point it is worth observing that in any attempt to go beyond formal considerations and rigorously prove $\lim_{\hbar \to 0}[\hat{P}_{q\alpha},\hat{P}_{r\beta}]=0$ for any indices $\alpha$ and $\beta$, the mathematical properties of the operators $\hat{P}_{q\alpha}$ and $\hat{P}_{r\beta}$ may play a crucial role. Thus, only particular choices of the projection operators $\hat{P}_{q\alpha}$ and $\hat{P}_{r\beta}$ might be suitable for rigorous arguments concerning the emergence of the logical bivalent limit.\\

\noindent Then again, one can imagine that a bivalent semantic only nearly emerges at the end of the measurement process when the size of the combined system, which includes the quantum interference experiment, the detectors, and the entire macroscopic environment, becomes infinitely large. In this way, the logical bivalent limit could be an idealization reserved for the limit where the size of the system is infinite, which can be symbolically denoted by ``the limit $N \to \infty$''.\\

\noindent However, as it can be shown (see, for example, \cite{Landsman}), such a limit is just a special case of the limit $\hbar \to 0$. That is, mathematically speaking, the nature of the idealization involved in assuming that a system is infinitely large is much the same as that of assuming $\hbar \to 0$ in a quantum system of a finite size. Accordingly, the observation about the difficulties of proving that the commutator $[\hat{P}_{q\alpha},\hat{P}_{r\beta}]$ comes to be 0 as $\hbar$ approaches 0 regards the limit $N \to \infty$ as well.\\

\section{Introducing supervaluationism in quantum mechanics}\label{Conclusion}  

\noindent It must be noted, however, that none of the offered above objections to the equality (\ref{17}) can be considered decisive.\\

\noindent Even so, a way to avoid at least some of those objections might be in the acknowledgment that in quantum mechanics truth values of the future non-certain events simply do not exist.\\

\noindent Such could be if any lattice element different from the least and the greatest elements carries no truth values, that is,\smallskip

\begin{equation} \label{19} 
   \left\{
   v\left(\ast\right) |  
   \, \ast \neq 0_L
   \;\: \text{and}
   \;\: \ast \neq 1_L
   \right\}
   =
   \emptyset
   \quad
   \text{whereas}
   \quad
   v(0_L)=0
   \quad
   \text{and}
   \quad
   v(1_L)=1
   \;\;\;\;  .
\end{equation}
\smallskip

\noindent Along the lines of this assumption, prior to the verification, an object $S=\{\diamond\}$ of a formal language to which an element of $L(\mathcal{H})$ other than the $\{0\}$-subspace and the identical subspace $\mathcal{H}$ is attributed, should not be called a proposition -- i.e., a primary bearer of truth-value -- but a propositional formula or a sentence or \textit{anything that carries no intrinsic meaning of truth or falsehood}.\\ 

\noindent Only subsequent to the act of verification, i.e., when the image of $\diamond$ under the valuation comes to be either 1 or 0, one can call the aforesaid object a proposition.
$\,$\footnote{\label{f8}This is reminiscent of the logical system of intuitionistic logic that lacks a complete set of truth values because its semantics is specified in terms of \textit{provability conditions}.\vspace{5pt}}
\\

\noindent Let's take, for example, a system whose state before the verification is given by a quantum superposition of the type $|{\Psi}\rangle = c_1|{\Psi}_{x1}\rangle + c_2|{\Psi}_{x2}\rangle$ such that $|{\Psi}_{x1}\rangle \in \mathrm{ran}(\hat{P}_{x1})$ and $|{\Psi}_{x2}\rangle \in \mathrm{ran}(\hat{P}_{x2})$ where $\mathrm{ran}(\hat{P}_{x1})$ and $\mathrm{ran}(\hat{P}_{x2})$ are closed subspaces of $\mathcal{H}$ that have no element in common except $\{0\}$, and $c_1$, $c_2$ are the superposition coefficients. The subspaces $\mathrm{ran}(\hat{P}_{x1})$ and $\mathrm{ran}(\hat{P}_{x2})$ are identified with simultaneously testable but disjoint (i.e., mutually exclusive) properties of the system, possession of which are declared by the propositional formulas $X_1$ and $X_2$ associated with the orthogonal projection operators  $\hat{P}_{x1}$ and  $\hat{P}_{x2}$.\\

\noindent It is straightforward that the superposition $c_1|{\Psi}_{x1}\rangle + c_2|{\Psi}_{x2}\rangle$ would correspond to the direct sum of $\mathrm{ran}(\hat{P}_{x1})$ and $\mathrm{ran}(\hat{P}_{x2})$ bringing in a decomposition of $\mathcal{H}$, namely,\smallskip

\begin{equation} \label{20} 
   \mathcal{H}
   =
   \mathrm{ran}(\hat{P}_{x1})
   \oplus
   \mathrm{ran}(\hat{P}_{x2})
   =
   \left\{
      c_1|{\Psi}_{x1}\rangle
      +
      c_2|{\Psi}_{x2}\rangle
   \right\}
   \;\;\;\;  .
\end{equation}
\smallskip

\noindent On the other hand, given that the meet of the orthogonal projections $\hat{P}_{xi}$ is ${\sqcap}_{i=1}^{2} \hat{P}_{xi} =\hat{P}_{x1}\hat{P}_{x2}=\hat{P}_{x2}\hat{P}_{x1} = 0$, the direct sum (\ref{20}) would correspond to the projection $\hat{P}_{x1} + \hat{P}_{x2}$ and, hence, the join ${\sqcup}_{i=1}^{2} \hat{P}_{xi}$.\\

\noindent In accordance with the definition (\ref{14}), one gets then\smallskip

\begin{equation} \label{21} 
   v\left(
      \mathrm{ran}(\hat{P}_{x1}) \cap \mathrm{ran}(\hat{P}_{x2})
   \right)
   =
   v\left(
      {\sqcap}_{i=1}^{2} \hat{P}_{xi}
   \right)
   =
   [\![X_1 \! \land \! X_2]\!]_v
   \;\;\;\;  ,
\end{equation}

\begin{equation} \label{22} 
   v\big(
      \left\{
         c_1|{\Psi}_{x1}\rangle +c_2|{\Psi}_{x2}\rangle
      \right\}
   \big)
   =
   v\left(
      {\sqcup}_{i=1}^{2} \hat{P}_{xi}
   \right)
   =
   [\![X_1 \! \lor \! X_2]\!]_v  
   \;\;\;\;  .
\end{equation}
\smallskip

\noindent Consistent with the assumption (\ref{19}), $\{v(\hat{P}_{xi})\}=\emptyset$ at the same time as $v(\mathrm{ran}(\hat{P}_{x1}) \cap \mathrm{ran}(\hat{P}_{x2})) =v(\{0\})= 0$ and $v(\{c_1|{\Psi}_{x1}\rangle +c_2|{\Psi}_{x2}\rangle\})=v(\mathcal{H})=1$, which would give\smallskip

\begin{equation} \label{23} 
   \left\{
   [\![X_i]\!]_v  
   \right\}
   =
   \emptyset
   \;\;\;\;  ,
\end{equation}

\begin{equation} \label{24} 
   [\![X_1 \! \land \! X_2]\!]_v 
   =
   0
   \;\;\;\;  ,
\end{equation}

\begin{equation} \label{25} 
   [\![X_1 \! \lor \! X_2]\!]_v  
   =
   1
   \;\;\;\;  .
\end{equation}
\smallskip

\noindent Accordingly, ahead of the verification, the sentence ``\textit{Out of two contradictory properties, the system possesses one or the other, but not both}'' would be true (and thus it would be a proposition) despite the fact that the sentence ``\textit{Out of two contradictory properties, the system possesses a particular one}'' would have no truth value at all (implying that before the verification the truth degree functions of the logical connectives $F_{\neg} ({[\![X_i]\!]}_v)$, $F_{\land} ({[\![X_1]\!]}_v, {[\![X_2]\!]}_v)$ and $F_{\lor} ({[\![X_1]\!]}_v, {[\![X_2]\!]}_v)$ would not be defined on ${[\![X_i]\!]}_v$).
$\,$\footnote{\label{f13}This inference concurs with the conclusion drawn in the paper \cite{Kochen} arguing that the major transformation from classical to quantum physics lies in the shift from intrinsic to extrinsic properties. In consequence, a compound property such as $X \!\lor\! Y$ may have a truth value, even though neither $X$ nor $Y$ has one.\vspace{5pt}}\\

\noindent Nevertheless, provided a probability assignment for the latter sentence is possible in a way that\smallskip

\begin{equation} \label{26} 
   \left\{
   [\![X_i]\!]_v  
   \right\}
   =
   \emptyset
   \implies
   0 < \mathbb{P}[X_i] < 1
   \;\;\;\;  ,
\end{equation}
\smallskip

\noindent the probability that this sentence will be proved to be true given the particular property $i \in \{1,2\}$ can be estimated by $\mathbb{P}[X_i] = \langle\Psi|\hat{P}_{xi}|\Psi\rangle = {|c_i|}^2$ where $|c_1 |^2+|c_2 |^2=1$.\\

\noindent As a rule, $\mathcal{H} \neq (\{c_1|{\Psi}_{x1}\rangle +c_2|{\Psi}_{x2}\rangle\}$ and so in general $\{[\![X_1 \! \lor \! X_2]\!]_v\} = \emptyset$ along with $\{[\![X_1 \! \land \! X_2]\!]_v\} = \emptyset$. This implies that prior to the verification the law of alternatives, i.e., $\mathbb{P}[X_1 \lor X_2] = \mathbb{P}[X_1] +\mathbb{P}[X_2]$, would not be valid on the whole.\\

\noindent So, essentially, the supervaluationist assumption (\ref{19}) suggests that in the scope of orthodox quantum mechanics and a related quantum logic, one should -- to paraphrase the remark made in the paper \cite{Sorkin} -- focus on maps from the family $L(\mathcal{H})$ to the unit interval $[0,1]$ that generalize the classical idea of probability, \textit{rather than that of truth}.

\section*{Acknowledgment}
\noindent The author would like to thank the anonymous referee for the inspiring feedback and the insights.\\

\bibliographystyle{References}
\bibliography{References}

\end{document}